\def\be{\begin{equation}}
\def\ee{\end{equation}}
\def\J{$J/\psi$}
\def\j{J/\psi}
\def\P{$\psi'$}
\def\p{\psi'}
\def\U{$\Upsilon$}
\def\u{\Upsilon}
\def\c{c{\bar c}}
\def\b{b{\bar b}}
\def\F{$\Phi$}
\def\f{\Phi}
\def\d{$d{\hat{\sigma}}$}

\def\d{\lambda}

\def\Q{Q{\bar Q}}

\def\l{\lambda}
\def\e{\epsilon}

\def\NP{{ Nucl.\ Phys.\ }}
\def\PL{{ Phys.\ Lett.\ }}
\def\PR{{ Phys.\ Rev.\ }}

\def\PRL{{ Phys.\ Rev.\ Lett.\ }}

\def\ZP{{ Z.\ Phys.\ }}

\def\la{\Lambda_{\rm QCD}}
\def\l{\lambda}

\def\Q{$Q{\bar Q}$}

\def\la{\Lambda_{\rm QCD}}
\def\F{$\Phi$}

\documentstyle[12pt,a4]{article}
\begin{document}
\thispagestyle{empty}
%\nopagenumbers
%\pageno=0
%Draft \hfill 18.5.1995\par
%\input macro-hs
\begin{flushright}
CERN-TH/95-342\\
BI-TP 95/41
\end{flushright}
\vskip1cm
\centerline{\Large{\bf Quarkonium Interactions in QCD\footnote{Presented 
at the Enrico Fermi International School of Physics on 
``Selected Topics in \\  
Non-Perturbative QCD", Varenna, Italy, June 1995; to appear 
in the Proceedings.}}}
\vskip0.6cm
\centerline{D. KHARZEEV}
\vskip0.3cm
\centerline{{\it Theory Division, CERN, CH-1211 Geneva, Switzerland}}
\par
\centerline{{\it and}}
\par
\centerline{{\it Fakult\"at f\"ur Physik, Universit\"at Bielefeld,
D-33501 Bielefeld, Germany}}
\vskip1cm
\centerline{CONTENTS}
\vskip0.2cm
\leftskip=1.5truecm
\begin{enumerate}
\item{Introduction \hfill \break
          1.1 Preview \hfill\break
          1.2 QCD atoms in external fields}
%\item{Quarkonium Production in Hadron-Hadron Collisions\hfill\break
%          2.1 Colour evaporation \hfill\break
%          2.2 Quarkonium production: Theory and Data}
\item{Operator Product Expansion for Quarkonium Interactions\hfill\break
          2.1 General idea\hfill\break
          2.2 Wilson coefficients
          \hfill\break
          2.3 Sum rules
          \hfill\break
          2.4 Absorption cross sections}
\item{Scale Anomaly, Chiral Symmetry and Low-Energy Theorems\hfill\break
          3.1 Scale anomaly and quarkonium interactions \hfill\break
          3.2 Low energy theorem for quarkonium interactions with pions
          \hfill\break
          3.3 The phase of the scattering amplitude}
\item{Quarkonium Interactions in Matter
          \hfill\break
          4.1 Nuclear matter
          \hfill\break
          4.2 Hadron gas
          \hfill\break
          4.3 Deconfined matter}
%\item{Confronting Experiment \hfill\break
%          5.1 How to measure low-energy quarkonium-nucleon cross section
%          \hfill\break
%          5.2 Quarkonium suppression as a signature of QGP revisited
%          \hfill\break
%          5.3 Is there an evidence for QGP from the present data?}
\item{Conclusions and Outlook}
\end{enumerate}
\bigskip\vfill
\leftskip=0truecm
\eject
\section{Introduction}

\subsection{Preview}
\medskip
\medskip\noindent
Heavy quarkonia have proved to be extremely useful for understanding QCD. 
The large mass of heavy quarks allows a perturbation theory
 analysis 
of quarkonium decays \cite{dec} (see \cite{Ger} for a recent review). 
Perturbation theory also provides a reasonable 
first approximation to the correlation functions of quarkonium currents; 
deviations from the predictions of perturbation theory can therefore 
be used to infer an information about the nature of non-perturbative 
effects. 
This program was first realized at the end of the seventies \cite{SVZ}; 
it turned out to be 
 one of the first steps towards a quantitative understanding of the QCD 
vacuum. 
 
The natural next step is to use heavy quarkonia to probe the properties of 
excited QCD vacuum, which may be produced in relativistic heavy ion 
collisions; this was proposed a decade ago \cite{MS}. 
This suggestion was based on the concept 
of colour screening of static potential acting between the heavy quark 
and antiquark. After a \J\ suppression was observed experimentally \cite{NA38}, 
 alternative ``conventional" explanations of the effect were proposed 
\cite{Blaizot}. 

The resulting ambiguity of the situation clearly calls for 
a more detailed analysis. An essential ingredient of this analysis 
has to be a dynamical treatment of quarkonium interactions with external 
gluon fields. This dynamical approach should then allow us to consider 
quarkonium interactions with vacuum gluon fields, gluon fields 
confined inside light hadrons and deconfined gluons on the same footing. 
This review is aimed at the description of recent progress in this direction.

The interactions of quarkonium 
with gluons are similar to the interactions of a hydrogen atom 
with external electromagnetic fields. 
This makes the problem interesting in itself from the pedagogical point of 
view, and allows us to draw a 
close analogy between the operator product expansion series of QCD and 
the conventional multipole expansion of atomic physics. We shall try to 
forward this analogy as far as possible, since it appeals to one's  
physical intuition and makes the discussion more clear and transparent. 

Let us briefly preview the things to come. 
Many conventional   
explanations of \J\ suppression are based on the crucial assumption 
that the absorption 
cross section of \J\ is equal to its geometrical value 
even at low energy. Though this kind of assumption is generally justified for 
light hadrons, we shall argue that this is not so for heavy quark--antiquark 
states. 

We shall discuss  
how the QCD theorems can be used to derive 
model- independent results for the amplitudes of quarkonium interactions 
with light hadrons at low energies.
We show how spontaneously broken scale and chiral symmetries imply a 
decoupling 
of Goldstone bosons from  
heavy and tightly bound quarkonium states. We demonstrate how the phase of the 
forward scattering amplitude of quarkonium interaction can be calculated 
directly from QCD with the use of dispersion relations and low-energy theorems. 

Finally we discuss quarkonium interactions with various kinds of 
QCD matter: nuclear matter, hadron gas and 
deconfined matter. We shall stress that the hardness of 
deconfined gluons  
opposed to the softness of gluon distributions in a confined medium  
can provide a clear-cut test of the state of QCD matter.

\subsection{QCD atoms in external fields}

The size of a bound state of a sufficiently heavy quark--antiquark pair 
is small enough to allow for a systematic QCD analysis of  
its interactions. Since there are no heavy (c,b,..) quarks inside light 
hadrons, the interaction of quarkonium with light hadrons is always 
mediated by gluon fields (which, at some distance larger than the quarkonium 
radius, couple to light quarks). This feature makes the problem of 
the interactions of heavy quarkonium somewhat similar to the problem 
of a hydrogen atom interacting with an external photon field. 
In this latter, much easier, problem one deals in particular with two 
different effects:

 i) when the external field is soft, it can change 
the static properties of the atom -- for instance its binding energy 
(a particular example of this is the Stark effect); 

ii) when the external 
photons are hard enough, they can break up the atom (photo--effect). 
In both cases the multipole expansion proves to be useful in calculating 
the level shifts and the transition probabilities.

A similar analysis has been carried out for the interaction of heavy 
quarkonium. It was demonstrated \cite{Got}--\cite{Leut} that the interaction 
of heavy quarkonium with the vacuum gluon field 
(``gluon condensate" \cite{SVZ}) leads to an analog of quadratic (due 
to the condition of colour neutrality) Stark effect, and this affects 
the properties of the $\bar{Q}Q$ state -- its mass, width, and the 
wave function. (The case of the interaction with the gluon fields
 characterized 
by a finite correlation time was considered in refs. \cite{time}.)   
%In a similar fashion one can evaluate the (real) low--energy 
%amplitude of quarkonium--hadron interaction \cite{Peskin,Bhanot,Luke,KV}. 
On the other hand, the dissociation of quarkonium  
in hadronic interactions can be viewed in complete analogy 
with the photo-effect \cite{Peskin}--\cite{KS} (see also \cite{ES}). 
In this case the dissociation can 
only occur if the gluon from the light hadron wave function is hard in 
the rest frame of the $\bar{Q}Q$ state, i.e. its energy is high enough 
to overcome the binding energy threshold. Since the gluon distributions 
inside light hadrons are generally soft, and the binding energy of heavy 
quarkonium is large, the condition of the gluon hardness is satisfied only 
when the relative momentum of the quarkonium and light hadron is very high. 
As a consequence of this, the absorption cross section of quarkonium rises 
very slowly from the threshold, reaching its geometrical value only at 
very high energy. When the gluon distributions in matter become {\it hard}, 
the behaviour of the absorption cross section changes drastically -- absorption 
becomes strong already at small energy. We will show later 
that this is very important for the diagnostics of {\it deconfined} matter 
\cite{KS,HS}.

\section{Operator Product Expansion for} 
\hskip1.3cm{\Large{\bf Quarkonium Interactions}}
\renewcommand{\theequation}{2.\arabic{equation}}
\setcounter{equation}{0}
\medskip
\subsection{General idea}

The QCD analysis of quarkonium interactions applies
to heavy and strongly bound quark--antiquark
states \cite{Bhanot}; therefore we restrict ourselves here to the
lowest $\c$ and $\b$ vector states \J~and \U, which we denote
generically by \F, following the notation of \cite{Bhanot}.
For such states, both the masses $m_Q$ of the constituent
quarks and the binding energies $\e_0(\f) \simeq (2M_{(Qq)}
- M_{\f}$) are much larger
than the typical scale $\Lambda_{\rm QCD}$ for non-perturbative
interactions; here $(Qq)$ denotes the lowest open charm or beauty
state. In $\f-h$ interactions, as well as in \F-photoproduction,
$\gamma h \to \f h$, we thus only probe a
small spatial region of the light hadron $h$; these processes are much
like deep-inelastic lepton--hadron scattering, with large $m_Q$ and
$\e_0$ in place of the large virtual photon mass $\sqrt{-q^2}$.
As a result, the calculation of \F-photoproduction and of
absorptive $\f-h$ interactions
can be carried out in the short-distance formalism
of QCD. Just like deep-inelastic leptoproduction, these reactions
probe the parton structure
of the light hadron, and so the corresponding cross sections can
be calculated in terms of parton interactions and structure
functions.
\par

Consider the amplitude for forward scattering of a virtual
photon on a nucleon,
\be
F(s, q^2) \sim
i\int d^4x e^{iqx} \langle N|T\{J_{\mu}(x)J_{\nu}(0)\}|N \rangle. \label{2.1}
\ee
In the now standard application of QCD to
deep-inelastic scattering one exploits the fact that at large
space-like photon momenta $q$ the amplitude is dominated by small
invariant distances of order $1/\sqrt{-q^2}$.
The Wilson operator product expansion then
allows the evaluation of the amplitude at the unphysical point
$pq\to 0$, where $p$ is the four-momentum of the nucleon.
Since the imaginary part of the amplitude (\ref{2.1}) is proportional
to the experimentally observed structure functions of deep-inelastic
scattering, the use of dispersion relations relates the value
of the amplitude at $pq\to 0$ point to the integrals over the
structure functions, leading to a set of dispersion sum rules
\cite{Gross}.
The parton model can then be considered as a particularly useful
approach satisfying these sum rules.
\par
In the case $J_{\mu} = \bar{Q} \gamma_{\mu} Q$, i.e. when
the vector electromagnetic current in Eq.\ (\ref{2.1}) is that
of a heavy quark--antiquark pair, large momenta $q$ are not needed
to justify the use of perturbative methods.
Even if $q\sim 0$, the small space-time scale of $x$ is set by the
mass of the charmed quark, and the characteristic distances
which are important in the correlator (\ref{2.1}) are of the order of
$1 /2m_Q$. In \cite{SVZ1,Novikov}, this observation was
used to
derive sum rules for charm photoproduction in a manner quite similar to
that used for deep-inelastic scattering.

In the interaction of quarkonium with light hadrons, again the
small space scale is set by the
mass of the heavy quark, and the characteristic distances involved
are of the order of quarkonium size, i.e. smaller than the
non-perturbative hadronic scale $\Lambda_{{\rm QCD}}^{-1}$.
Also, since heavy quarkonium and light hadrons do not have
quarks in common, the only allowed exchanges are purely gluonic.
However, the smallness of spatial size is not enough to justify the use
of perturbative expansion \cite{Bhanot}. Unlike in the case of
\F-photoproduction, heavy quark lines now appear in the initial
and final states, so that the \Q~state can emit and
absorb gluons at points along its world line widely separated in
time. These gluons must be hard enough to interact
with a compact colour singlet state (colour screening leads to a
decoupling of soft gluons with wavelengths larger than the
size of the \F); however, the interactions among the gluons
can be soft and non-perturbative. We thus have to assure that the
process is well localized also in time.
Since the absorption or emission of a gluon turns
a colour singlet quarkonium state into a colour octet,
the scale that regularizes the time correlation of such processes
is just, by way of the quantum-mechanical uncertainty
principle, the mass difference between the colour-octet and 
colour-singlet states of quarkonium: $t_{Q} \sim 1/ (\e_8 - \e_1)$.
The perturbative Coulomb-like piece of the heavy quark--antiquark
interaction
\be
V_k(r)=-g^2 {c_k \over {4\pi r}} \label{2.2}
\ee
is attractive in the colour singlet ($k=1$) and repulsive in the
colour-octet ($k=8$) state; in SU($N$) gauge theory
\be
V_s=-{g^2 \over {8\pi r}} {N^2-1 \over N}, \label{2.3a}
\ee
\be
V_a={g^2 \over {8\pi r}} {1 \over N}. \label{2.3b}
\ee
To leading order in $1/N$, the mass gap between the singlet and adjoint 
states is therefore just the binding energy of the heavy
quarkonium $\e_0$, and the characteristic correlation time for
gluon absorption and emission is
\be
t_Q \sim 1/\e_0. \label{2.4}
\ee
It is important to note that, for Coulombic binding, the mass gap between 
the singlet and adjoint states is given by 
\be
\e_a - \e_s = \left\langle {g^2 \over {8\pi r}} \right\rangle \ N,
\ee
i.e. it increases with $N$, and the lifetime of the adjoint state 
becomes very small in the large-$N$ limit . The mean size of the state also 
decreases with $N$, since the attractive potential (\ref{2.3a}) is 
proportional 
to it. This implies that the operator product expansion is applicable 
 in the large-$N$ limit even when the heavy quark mass is not too large. 
Also, the interaction between the quarks in the adjoint state (\ref{2.3b}) 
is suppressed by the factor $1/N$. This allows us to neglect the final-state 
interaction in considering the dissociation of quarkonium in the large-$N$ 
limit. This approximation  
is also applicable for $N=3$ since its accuracy is about $1/N^2$. 
(It is well known that 
the final-state interactions are important in photo-effect on 
hydrogen atoms.)
\vskip0.3cm

For sufficiently heavy quarks, the dissociation of quarkonium states
by interaction with light hadrons can thus be fully accounted for
by short-distance QCD. Such perturbative calculations become valid
when the space and time scales associated with the quarkonium state,
$r_Q$ and $t_Q$, are small in comparison with the non-perturbative
scale $\la^{-1}$
\be
r_Q<<\la^{-1}, \label{2.26a}
\ee
\be
t_Q<<\la^{-1}; \label{2.26b}
\ee
$\la^{-1}$ is also the characteristic size of light hadrons.
In the heavy quark limit, the quarkonium binding becomes Coulombic,
and the spatial size $r_Q \sim (\alpha_sm_Q)^{-1}$ is thus small.
The time scale is given by the uncertainty relation as the inverse
of the binding energy $E_Q \sim \alpha_sm_Q$ and hence also small.

For the charmonium ground state $\j$, we have
\be
r_{\j} \simeq 0.2~{\rm fm}= (1~ {\rm GeV})^{-1};~~~
E_{\j} = 2 M_{D} - M_{\p} \simeq 0.64~{\rm GeV}. \label{2.27}
\ee
With $\la \simeq 0.2$ GeV, the inequalities (\ref{2.27}) seem already
reasonably well satisfied, and also the heavy quark relation
$E_{\j} = (1/m_c r_{\j}^2)$ is very well fulfilled.
 For the \U, the interaction
is in fact essentially Coulomb-like and the mass gap to open beauty
is even larger than for charm. One therefore expects to be able to treat
quarkonium interactions with light hadrons by the same QCD methods
that are used in deep-inelastic scattering and charm photoproduction.
\vskip0.3cm

An important feature of the quarkonium structure is the small velocity 
of heavy quarks inside it: $v \sim \alpha_s$. This simplifies the 
calculations, since in the non-relativistic domain one can keep only 
the chromo-electric part of the interaction with the external gluon fields -- 
the chromo-magnetic 
part will be suppressed by higher powers of velocity. This reduces the 
number of terms in the OPE series and makes the entire calculation 
more reliable: most of the results become exact in the heavy 
quark limit. 
Since even the charm quark may be sufficiently heavy to ensure relatively 
large binding energy and small velocity of constituents, we expect that 
the operator product expansion can provide a reasonable description already 
for the interactions of \J's. 
\par

\subsection{Wilson coefficients}

We shall use the operator product
expansion to compute the amplitude of heavy quarkonium
interaction with light hadrons,
\be
F_{\f h} = i\int d^4x e^{iqx} \langle h|T\{J(x)J(0)\}|h \rangle =
\sum_n c_n(Q,m_Q) \langle O_n \rangle ,
\label{2.5}
\ee
where the set $\{O_n\}$ should include all local gauge-invariant operators
expressible in terms of gluon fields; the matrix elements $\langle
O_n \rangle$ are taken between the initial and final light-hadron
states. The coefficients $c_n$ are expected to be computable perturbatively
and are process-independent. 

The case of quarkonium interactions with hadrons provides a particularly 
transparent 
illustration of the operator product expansion scheme:  
the OPE series has a structure which 
recalls the usual multipole expansion for an atom in the external 
electromagnetic field. Indeed, the amplitude in the rest frame of 
quarkonium can be written in the following simple form:
\be
F_{\f h} = {g^2 \over 2 N} \left\langle\ {\vec r} {\vec E}^a 
{1 \over H_{a} + \epsilon 
+ i D^0} {\vec r} {\vec E}^a\ \right\rangle, \label{2.6}
\ee
where ${\vec E}$ is the chromo-electric field, $D^0$ is the covariant 
derivative, and $H_a$ is the Hamiltonian of the colour-octet $\bar{Q}Q$ state. 
It is evident that (\ref{2.6}) is just the amplitude corresponding to 
the quadratic Stark effect in the external gluon field. We can expand the 
denominator in (\ref{2.6}) representing the amplitude as a series containing 
matrix elements of gauge-invariant operators:
\be
F_{\f h} = {g^2 \over 2 N} \sum_{n=2}^{\infty} 
\langle \f | r^i {1 \over{(H_{a} + \epsilon )^{n-1}}} r^j | \f \rangle 
\langle h|E^a_i(D^0)^{n-2}E^a_j|h \rangle ,\label{2.7}
\ee
where the sum runs over the even values of $n$.
To eliminate explicit dependence of (\ref{2.7}) on the 
coupling $g$ and the number 
of colours $N$, one can express it in terms of the Bohr radius $r_0$ and 
Rydberg energy $\epsilon_0$:
\be
F_{\f h} = r_0^3 \epsilon_0^2 \sum_{n=2}^{\infty} d_n 
 \langle h|{1 \over 2} G^a_{0i}(D^0)^{n-2}G^a_{0i}|h \rangle ,\label{2.8c}
\ee  
where we have introduced the gluon field operators $G^a_{0i} = - E^a_i$. 
The expression (\ref{2.8c}) is manifestly gauge-invariant and realizes the 
operator product expansion (\ref{2.5}) in terms of twist-two gluon field 
operators. The dimensionless parameters 
$d_n$ in the sense of the OPE (see (\ref{2.5})) are the Wilson coefficients, 
which 
are defined as
\be
d_n= {36 \over N^2} r_0^{-3}\  
\langle \f | r^i {1 \over{(H_{a} + \epsilon )^{n-1}}} r^i | \f \rangle . 
\ee
For $1S$ and $2S$ states these coefficients were computed in \cite{Peskin} 
in the leading order in $1/N^2$:
\be
d_n^{(1S)}=\left({32 \over N}\right)^2 \sqrt{\pi}\ {\Gamma(n+{5\over 2}) \over 
\Gamma(n+5)}; \label{2.14a}
\ee
\be
d_n^{(2S)}=\left({32 \over N}\right)^2 4^n \sqrt{\pi}\ 
{\Gamma(n+{5\over 2}) \over \Gamma(n+7)} (16n^2 + 56n + 75).
\ee
For the sake of completeness, we give here also the expression for the 
$2P-$states of quarkonium:
\be
d_n^{(2P)}=\left({15 \over N}\right)^2 4^n 2 \sqrt{\pi}\  
{\Gamma(n+{7\over 2}) \over \Gamma(n+6)}.
\ee
As mentioned above, in
deep-inelastic scattering
the expansion (\ref{2.5}) is useful only in the vicinity of the point
$pq\to 0$. The same is true in our case, as we shall now discuss.
Indeed, the matrix element of any tensor operator $O_n^{\mu_1...\mu_k}$ 
in a hadron 
state with the momentum $p_{\mu}$, averaged over the hadron spin, 
has the form
\be
\langle h|O_n^{\mu_1...\mu_k}|h \rangle = p_{\mu_1}...p_{\mu_k} C_n^k\ -\ 
{\rm traces},
\ee
where $C_n^k$ are scalar ``irreducible" matrix elements which carry 
the information about the structure of the hadron. Since the only vector 
associated with a spin-averaged \F\ state is its momentum $q_{\mu}$, 
the matrix element of the tensor operator can appear in the expansion of the 
amplitude only in the form
\be
q_{\mu_1}...q_{\mu_k} \langle h|O_n^{\mu_1...\mu_k}|h \rangle = 
(pq)^n C_n^k, 
\ee
where we have omitted the contribution of terms containing traces; these 
lead to the target mass corrections and can be systematically taken 
into account. 

Applying these arguments to our case, we come to the following expression 
for the matrix element of gluon fields in a hadron:
\be
\langle h|{1 \over 2} G^a_{0i}(D^0)^{n-2}G^a_{0i}|h \rangle = 
\langle O_n \rangle \ \left({\lambda \over \epsilon_0}\right)^n, 
\ee
where we have introduced the variable
\be
\lambda = {pq \over M_{\f}} = {(s-M_{\f}^2 - M_h^2)\over 2M_{\f}}
\simeq {(s-M_{\f}^2)\over 2M_{\f}}.
\label{2.6a}
\ee
In the rest frame of quarkonium, 
$\lambda$ is the energy of the incident hadron. 
The approximate equality in (\ref{2.6a}) becomes valid in the heavy quark 
limit, when one can 
neglect the mass of the light hadron $M_h$.  The dimensionless scalar 
matrix elements $\langle O_n \rangle$ carry the information about the 
gluon fields inside the light hadron.
We therefore obtain
\be
F_{\f h} = r_0^3\ \e_0^2\ \sum_{n=2}^{\infty}
d_n \langle O_n \rangle
\left({\lambda \over \e_0}\right)^n, \label{2.7a}
\ee
where the sum runs over even values of $n$; this ensures the
crossing symmetry of the amplitude.
\par

\subsection{Sum rules}

Since the total $\f-h$ cross section $\sigma_{\f h}$ is proportional
to the imaginary part of the amplitude $F_{\f h}$, the dispersion
integral over $\l$ leads to the sum rules
\be
{2\over \pi}\int^{\infty}_{\l_0} d\l~\l^{-n} \sigma_{\f h}(\l)
= r_0^3~ \e_0^2~ d_n \langle O_n \rangle \left( {1\over \e_0}
\right)^n.
\label{2.8}
\ee
Equation\ (\ref{2.8}) provides only the inelastic intermediate states in the
unitarity
relation, since direct elastic scattering leads to contributions of
order $r_0^6$. Hence the total cross section in Eq.\ (\ref{2.8}) is due to
absorptive interactions only \cite{Bhanot}, and the integration
in Eq.\ (\ref{2.8}) starts at the lower limit $\l_0 > M_h$.
 Recalling now the expressions for radius and binding energy of
1S Coulomb bound states of a heavy quark--antiquark pair,
\be
r_0 = \left( {16\pi \over {3 g^2}} \right) {1 \over m_Q}, \label{2.9}
\ee
\be
\e_0 = \left( {3 g^2 \over {16 \pi}} \right)^2 m_Q, \label{2.10}
\ee
and using the coefficients $d_n$ from
(\ref{2.14a}), it is possible \cite{Bhanot} to rewrite these sum rules
in the form
\be
\int_{\l_0}^{\infty} { d\lambda \over \l_0} \left( {\lambda \over
\l_0} \right)^{-n}
\sigma_{\f h} (\lambda) = 2 \pi^{3/2}\ \left(16 \over 3 \right)^2\
{{\Gamma \left(n + {5\over 2} \right)} \over {\Gamma (n + 5)}}\
\left( {16\pi \over {3g^2}}\right)\
{1 \over m_Q^2}\ \langle O_n \rangle, \label{2.11}
\ee
with $\l_0/\e_0 \simeq 1$ in the heavy-quark limit.
The contents of these sum rules become more transparent
in terms of the parton model. In parton language,
the expectation values $\langle O_n \rangle$ of the operators composed
of gluon fields can be expressed as Mellin
transforms \cite{Parisi} of the gluon structure function of the light
hadron, evaluated at the scale $Q^2=\e_0^2$,
\be
\langle O_n \rangle = \int_0^1 dx\ x^{n-2} g(x, Q^2 = \e_0^2).
\label{2.12}
\ee
Defining now
\be
y = {\l_0 \over \lambda} , \label{2.13}
\ee
we can reformulate Eq.\ (11) to obtain
\be
\int_0^1 dy\ y^{n-2} \sigma_{\f h}(\l_0 / y) = I(n)\
\int_0^1 dx\ x^{n-2} g(x, Q^2 = \e_0^2), \label{2.14}
\ee
with $I(n)$ given by
\be
I(n) =  2 \pi^{3/2}\ \left(16 \over 3 \right)^2\
{{\Gamma \left(n + {5\over 2} \right)} \over {\Gamma (n + 5)}}\
\left( {16\pi \over {3g^2}}\right)\ {1 \over m_Q^2}. \label{2.15}
\ee
Equation\ (\ref{2.14}) relates the $\f-h$ cross section to the gluon structure
function. 

\subsection{Absorption cross sections}

To get a first idea of this relation, we neglect the
$n$-dependence of $I(n)$ compared with that of $\langle O_n \rangle$; then
we conclude that
\be
\sigma_{\f h}(\l_0 / x) \sim g(x, Q^2=\e_0^2), \label{2.16}
\ee
since all-order Mellin transforms of these quantities are equal up to a
constant. From Eq. (\ref{2.16}) it is clear that
the energy dependence of the $\f-h$ cross section is entirely
determined by the $x$-dependence of the gluon structure function.
The small-$x$ behaviour of the structure function governs
the high energy form of the cross section, and the hard tail
of the gluon structure function for $x \to 1$ determines
the energy dependence of $\sigma_{\f h}$ close to the
threshold.

To obtain relation (\ref{2.16}), we have neglected the $n$-dependence of the
function $I(n)$. Let us now try to find a more accurate solution of the
sum rules (\ref{2.14}). We are primarily interested in the energy region not
very far from the inelastic threshold, i.e.
\be
(M_h + \e_0)\ < \lambda <\ 5\ {\rm GeV}, \label{2.17}
\ee
since we want to calculate in particular the absorption of \F's
in confined hadronic matter. In such an environment,
the constituents will be hadrons with momenta of at most
a GeV or two. A usual hadron ($\pi,~\rho$, nucleon) of 5 GeV momentum,
incident on a \J~at rest, leads to $\sqrt s \simeq 6$ GeV, and this
corresponds to \\ 
$\l \simeq 5$ GeV.

From what we learned above, the energy region corresponding to the range
(\ref{2.17}) will be determined by the gluon structure function at
values of $x$ not far from unity. There
the $x$-dependence of $g(x)$ can be well described by a power law
\be
g(x) = g_2\ (k+1)\ (1-x)^k, \label{2.18}
\ee
where the function (\ref{2.18}) is normalized so that the second moment (4.12)
gives the fraction $g_2$ of the light hadron momentum carried
by gluons, $ \langle O_2\rangle = g_2 \simeq 0.5$.
This suggests a solution of the type
\be
\sigma_{\f h}(y) = a (1-y)^{\alpha}, \label{2.19}
\ee
where $a$ and $\alpha$ are constants to be determined.
Substituting (\ref{2.18}) and (\ref{2.19}) into the sum rule (\ref{2.14}) 
and performing
the integrations, we find
\be
a\ {{\Gamma(\alpha+1)} \over {\Gamma(n+\alpha)}} =
\left({2 \pi^{3/2} g_2 \over m_Q^2}\right)\left(16 \over 3 \right)^2
\left( {16\pi \over {3g^2}}\right)
{\Gamma(n+{5 \over 2}) \over \Gamma(n+5)} {\Gamma(k+2) \over
\Gamma(k+n)}.
\label{2.20}
\ee
We are interested in the region of low to moderate
energies; this corresponds to relatively large $x$, to which higher
moments are particularly sensitive.
Hence for the
range of $n$ for which Eq.\ (\ref{2.20}) is valid,
$n \leq 8$, the essential
$n$-dependence is contained in the $\Gamma$-functions.
For $n~\geq~4$, Eq.\ (\ref{2.20}) can be
solved in closed form by using an appropriate approximation for the
$\Gamma$-functions. We thus obtain
\be
a\ {\Gamma(\alpha +1) \over \Gamma(k+2)} \simeq {\rm const} \times 
n^{\alpha - k -5/2}. \label{2.21}
\ee
Hence to satisfy the sum rules (\ref{2.14}), we need
\be
\alpha=k+{5 \over 2}~~~~~~~
a = {\rm const} \times {\Gamma(k+2) \over \Gamma(k+{7 \over 2})}. \label{2.22}
\ee
Therefore the solution of the sum rules (\ref{2.14}) for
moderate energies $\lambda$ takes the form
\be
\sigma_{\f h}(\lambda) = 2 \pi^{3/2} g_2  \left(16 \over 3 \right)^2
\left( {16\pi \over {3g^2}}\right){1 \over m_Q^2}
{\Gamma(k+2) \over \Gamma(k+{7 \over 2})}
\left(1-{\l_0 \over \l}\right)^{k+5/2}. \label{2.23}
\ee
To be specific, we now consider the \J--nucleon interaction.
Setting $k=4$ in accordance with quark counting rules, using
$g_2\simeq 0.5$ and expressing the strong coupling $g^2$ in terms of
the binding energy $\e_0$ (Eq.\ (\ref{2.10})),
we then get from Eq.\ (\ref{2.23}) the energy dependence of
the $\j$--$N$ total cross section
\be
\sigma_{\j N}(\l) \simeq 2.5\ {\rm mb} \times
\left(1-{\l_0 \over \l}\right)^{6.5}, \label{2.24}
\ee
with $\l$ given by Eq.\ (6) and $\l_0 \simeq (M_N+\e_0)$.
This cross section rises very slowly from threshold; 
for $P_N \simeq 5 $ GeV, it is around 0.1 mb, i.e. more than an
order of magnitude below its asymptotic value.
\par
We should note that the high energy cross section of 2.5 mb in Eq.\
(\ref{2.24}) is {\sl calculated} in the short-distance formalism of QCD and
determined numerically by the values of $m_c$ and $\e_0$.
 From Eqs.\ (\ref{2.10}) and (\ref{2.23}), it is seen to be proportional to
$1/(m_Q\sqrt{m_Q \e_0})$. For $\u-N$ interactions, with $m_b\simeq$ 4.5
GeV and\\ 
$\e_0 \simeq 1.10$ GeV, we thus have the same form (\ref{2.24}),
but with
\be
\sigma_{\u N} \simeq 0.37~{\rm mb} \label{2.25}
\ee
as high-energy value. This is somewhat smaller than that obtained from
geometric arguments \cite{Povh} and potential theory \cite{MT}.
\par
\newpage

\section{Scale Anomaly, Chiral Symmetry and}
\hskip1cm{\Large{\bf Low-Energy Theorems}}
\renewcommand{\theequation}{3.\arabic{equation}}
\setcounter{equation}{0}

\subsection{Scale anomaly and quarkonium interactions}

Let us consider the amplitude of the quarkonium--hadron 
interaction (\ref{2.8}) 
at low energy. Specifically, we will consider the case when the energy 
of the incident hadron $E_h$ in the rest frame of quarkonium is much smaller 
than its binding energy: $E_h << \epsilon_0$. It is easy to check that 
$E_h$ is just identical to the variable $\lambda$ introduced earlier (see 
(\ref{2.6a})), and in the domain where $\lambda/\epsilon_0<<1$ we can keep 
only the lowest power of $n=2$ in the expansion (\ref{2.8}). The low-energy 
amplitude then becomes
\be
F_{\f h} = r_0^3\  d_2\ \lambda^2\ \langle h|O_2|h \rangle. \label{3.1}
\ee  
It is not difficult to identify the operators $O_2$ in the quarkonium 
rest frame; just as in atomic physics, $O_2$ can contain either the 
square of (chromo-)electric fields (quadratic Stark effect) or 
(chromo-)magnetic fields (quadratic Zeeman effect):
\be
\lambda^2\ \langle h|O_2^E|h \rangle  = {1 \over 9} 
\langle h| g^2 {\vec E}^a {\vec E}^a|h 
\rangle, \label{3.2a}
\ee
\be 
\lambda^2\ \langle h|O_2^B|h \rangle  = {1 \over 9} 
\langle h| g^2 {\vec B}^a {\vec B}^a|h 
\rangle. \label{3.2}
\ee
The reader can check that (\ref{3.1}) and (\ref{3.2a}) reproduce the 
first term 
in the expansion of the amplitude (\ref{2.7a}) introduced in Sect.2.  
Equations (3.1) and (3.2) determine the low-energy amplitude of quarkonium 
interactions in terms of the Wilson coefficient $d_2$ (calculated already in 
Sect.2.2) and the strength of colour fields inside a hadron.
 
Surprisingly enough, the latter quantity is fixed by low-energy 
QCD theorems \cite{JE},\cite{scale} and can be evaluated in a 
model-independent way \cite{VZ,KV,LMS}. 
To see this, let us write down, following \cite{LMS}, the operators (3.2) 
as linear combinations of the twist-two gluon operator
\be
M_2^{\mu\nu}={1\over 4} g^{\mu\nu} G^{\alpha\beta a}G^a_{\alpha\beta} - 
G^{\mu\alpha a}G^{\nu a}_{\alpha}, \label{3.3}
\ee
and the ``anomalous" part of the trace of the energy--momentum tensor of QCD:
\be
T^{\alpha}_{\alpha}={\beta (g) \over 2g} G^{\alpha\beta a}G^a_{\alpha\beta}. 
\label{3.4}
\ee 
In an arbitrary frame, where the quarkonium state moves with the four-velocity 
$v_{\mu}$, the operators (\ref{3.2a}),(\ref{3.2}) can be written down as
\be
{\vec E}^a {\vec E}^a = M_2^{\mu\nu} v_{\mu} v_{\nu} - {g \over 2\beta (g)} 
T^{\alpha}_{\alpha}, \label{3.5a}
\ee
\be
{\vec B}^a {\vec B}^a = M_2^{\mu\nu} v_{\mu} v_{\nu} +  {g \over 2\beta (g)} 
T^{\alpha}_{\alpha}. \label{3.5b}
\ee 
The matrix element of the operator $M_2^{\mu\nu}$ in a hadron state at 
zero momentum transfer is already familiar to us from Sect.2.3 (see (2.12), 
(2.18));
it is proportional to the fraction $g_2$ of the hadron momentum carried 
by gluons at the scale $Q^2=\epsilon_0^2$:
\be
\langle h| M_2^{\mu\nu} |h \rangle = 2 g_2 \left(p^{\mu}p^{\nu} - {1\over 4} 
g^{\mu\nu} p^2\right). \label{3.6}
\ee
The matrix element of $T^{\alpha}_{\alpha}$ is relevant only for low-energy 
interactions and was not evaluated before.  
To evaluate it, 
we need to have a closer look at the properties of the energy--momentum 
tensor of QCD. The trace of this tensor is given by
\be
\Theta^{\alpha}_{\alpha} = {\beta (g) \over 2g} 
G^{\alpha\beta a}G^a_{\alpha\beta} + \sum_{l=u,d,s} m_l (1 + \gamma_{m_l})
\bar{q_l}q_l + 
\sum_{h=c,b,t} m_h (1 + \gamma_{m_h})\bar{Q_h}Q_h, \label{3.7}
\ee   
where $\gamma_{m}$ are the anomalous dimensions; in the following we
will assume that the current quark masses are redefined as $(1 + \gamma_m)m$.  
The QCD beta function at the scale $Q^2=\epsilon_0^2$ can be written as 
\be
\beta (g) = - b {g^3 \over 16\pi^2} + ..., \ b = 9 - {2 \over 3} n_h,
\label{3.8}
\ee
where $n_h$ is the number of heavy flavours ($c,b,..$). 
Since there is no valence heavy quarks inside light hadrons,  
one expects a decoupling of heavy flavours at the scales $Q^2<4m_h^2$. 
This decoupling was consistently treated in the framework of the 
heavy-quark expansion \cite{SVZ2}; to order $1/m_h$, 
only the triangle graph with 
external 
gluon lines contributes. Explicit calculation shows \cite{SVZ2} that 
the heavy-quark 
terms transform in the piece of the anomalous gluonic part of $\Theta^{\alpha}
_{\alpha}$:
\be
\sum_{h} m_h \bar{Q_h}Q_h \to -{2\over 3}\ n_h\ {g^2\over 32\pi^2}  
G^{\alpha\beta a}G^a_{\alpha\beta} + ... \label{3.9}
\ee
It is immediate to see from (3.9), (3.7) and (3.8) that the heavy-quark 
terms indeed cancel the part of anomalous gluonic term associated 
with heavy flavours, so that the matrix element of the energy--momentum 
tensor can be rewritten in the form
\be
\Theta^{\alpha}_{\alpha} = {\tilde{\beta} (g) \over 2g} 
G^{\alpha\beta a}G^a_{\alpha\beta} + \sum_{l=u,d,s} m_l
\bar{q_l}q_l, \label{3.10}
\ee 
where heavy quarks do not appear at all; the beta function in (3.10) 
includes the contributions of light flavours only:
\be
\tilde{\beta} (g) = - 9 {g^3 \over 16\pi^2} + ...
\label{3.11}
\ee

To complete the calculation of the quarkonium scattering amplitude, 
we need only to recall that the matrix element of the energy--momentum tensor 
in a hadron state $|h\rangle$ at zero momentum transfer is 
defined as\footnote{Throughout this paper 
we use a 
relativistic normalization of hadron states $\langle h|h \rangle = 2 M_h V$, 
where $V$ is a normalization volume.} 
\be
\langle h|\Theta^{\alpha}_{\alpha}|h \rangle = 2 M_h^2, \label{3.12}
\ee
where $M_h$ is the hadron mass.   
Consider first the chiral limit $m_l \to 0$. In this limit the quark terms 
in the energy--momentum tensor can be omitted, and only the anomalous term 
contributes. The matrix elements of the operators (3.5) therefore take
the following form:
\be
\langle h|{\vec E}^a {\vec E}^a|h \rangle = \langle h|M_2^{\mu\nu}|h \rangle 
v_{\mu} v_{\nu} + {4\pi \over 9 \alpha_s} M_h^2, \label{3.13a}
\ee
\be
\langle h|{\vec B}^a {\vec B}^a|h \rangle = \langle h|M_2^{\mu\nu}|h \rangle 
v_{\mu} v_{\nu} - {4\pi \over 9 \alpha_s} M_h^2, \label{3.13b}
\ee
where we have introduced $\alpha_s=g^2/4\pi$ at the scale $Q^2=\epsilon_0^2$. 
It is evident that at small velocity the second terms on the r.h.s. of 
(\ref{3.13a},\ref{3.13b}), which are proportional to $\alpha_s^{-1}$, 
dominate. 
Furthermore, 
for non-relativistic quarks inside the quarkonium, the chromo-magnetic  
interaction is suppressed with respect to the chromo-electric one by the 
square 
of quark velocity $v_Q^2 \sim (m_Q r_0)^{-2} \simeq \alpha_s^2 << 1$.  
The amplitude of quarkonium--hadron  
interactions at low energy thus takes the form
\be
F_{\f h} \simeq r_0^3\  d_2\ {2 \pi^2 \over 27} \left(2M_h^2 - 
\langle h| \sum_{l=u,d,s} m_l \bar{q_l}q_l |h \rangle \right). \label{3.14}
\ee
One can see from (3.14) that, for example, the amplitude of a 
quarkonium--proton 
interaction at low energy is completely determined by the proton mass $M_p$, 
the value of the pion--nucleon $\Sigma$-term 
\be
\Sigma_{\pi N} = {\hat m} \langle p|\bar{u}u + \bar{d}d|p \rangle, \label{3.15}
\ee
with $\hat{m} = 1/2(m_u + m_d)$ (we ignore isospin splitting), 
and the strangeness contents of the proton (we recall that 
$d_2$ is a c-number, which was calculated in Sect.2.2):
\be
F_{\f p} \simeq r_0^3\  d_2\ {2 \pi^2 \over 27} 2M_p ( M_p - 2 \Sigma_{\pi N} -
\langle p| m_s \bar{s}s |p \rangle ). \label{3.16}
\ee
The empirical value of the pion--nucleon $\Sigma$-term extracted from the 
low-energy $\pi N$ scattering amplitude, $\Sigma_{\pi N} = 49\pm 7$ MeV, 
suggests that it can safely be omitted in (\ref{3.16}). The relatively large 
mass of the strange quark and non-negligible admixture of strange quarks 
in the proton can make the corresponding term in (\ref{3.16}) important. 
Indeed, the analysis of \cite{GLS} implies that
\be 
\langle p| m_s \bar{s}s |p \rangle = {m_s \over 2 {\hat m}}\ y\ \Sigma_{\pi N} 
\simeq 13 \times 0.2 \times 49\ {\rm MeV} \simeq 127\ {\rm MeV}, \label{3.17}
\ee
where $y$ is the relative scalar density of strange quarks in the proton, 
$y=2\langle p|\bar{s}s|p \rangle / \langle p|\bar{u}u + \bar{d}d|p \rangle$.
However, in the case of interactions with protons, the chiral limit 
($m_u, m_d, m_s$ $\to$ $0$) 
is still a reasonable approximation.
In the combined limit of small velocity of quarkonium and massless $u,d,s$ 
quarks, we get 
 a particularly simple 
expression for the low-energy amplitude:
\be 
F_{\f p} \simeq r_0^3\  d_2\ {4 \pi^2 \over 27} M_p^2, \label{3.18}
\ee
which is completely determined by the wave function of quarkonium and the 
mass of the proton. With the value of the Wilson coefficient $d_2$ from 
(\ref{2.14a}) it coincides with the result of ref. \cite{KV}, but differs 
from the result of ref. \cite{LMS}.
As it follows from (\ref{3.18}), in the chiral limit there 
is no explicit scale 
dependence in the amplitude. This is a consequence of the scale 
independence of 
the ``anomalous" gluon piece (\ref{3.4}) of the energy--momentum tensor.

It is important to note that the amplitude (\ref{3.18}) is purely real; 
physically, this means that the processes of quarkonium dissociation 
in the kinematical domain considered here (the momenta of incident 
hadrons in the $\Phi$ rest frame are small compared with the binding energy), 
dissociation of quarkonium is not possible\footnote{Even though the 
rearrangement processes, as $J/\Psi + N \to \Lambda_c + D$, 
are kinematically 
allowed even at low energy, they are dynamically suppressed in the heavy 
quark limit.}. 

The sign of the amplitude corresponds to an attraction. This makes possible 
the existence of nuclear bound states of quarkonium (first discussed 
in ref.\cite{Bro}). Also, it implies 
that in a dense hadron gas the quarkonium binding energy will effectively 
increase.

\subsection{Low-energy theorem for quarkonium interactions}
\hskip1.4cm{\large{\bf with pions}}

\vskip0.3cm

Naive application of the formula (\ref{3.18}) to the interactions with 
pions yields 
an amplitude proportional to $M_{\pi}^2$. However this result is not 
consistent, since in deriving (\ref{3.18}) we have used the chiral limit of  
$m_u, m_d, m_s \to 0$, and chiral symmetry tells us that in this limit 
the pion should become a Goldstone boson with zero mass. The origin of 
the difficulty can be traced back to the expression (\ref{3.10}) 
for the energy--momentum tensor of QCD from which it may seem that, 
in the 
chiral limit, the mass of the pion does not vanish because of  
 the gluon contribution arising from the scale anomaly. 
We shall show in this section 
 that this is 
not true, and that 
the spontaneously broken chiral and scale symmetries imply decoupling of 
low-energy pions from heavy 
quarkonium.

Let us take the matrix element of the trace of the energy--momentum tensor 
in a pion state:
\be
2 M_{\pi}^2 = \langle \pi | {\tilde{\beta} (g) \over 2g} 
G^{\alpha\beta a}G^a_{\alpha\beta} | \pi \rangle + 
\langle \pi | \sum_{l=u,d,s} m_l \bar{q_l}q_l | \pi \rangle, \label{3.19}
\ee 
where in the l.h.s. we have used the definition (\ref{3.12}). 
Current algebra tells 
us that
\be
\langle \pi | \sum_{l=u,d,s} m_l \bar{q_l}q_l | \pi \rangle = 2 M_{\pi}^2. 
\label{3.20}
\ee
Substituting (\ref{3.20}) into (\ref{3.19}) we find that the 
matrix element of the 
operator containing gluon fields in a pion must be equal to zero! Since this 
latter matrix element enters the low-energy amplitude of quarkonium--pion  
scattering, this implies decoupling of soft pions from heavy and tightly 
bound quarkonium. 

This result has a deep physical origin. Indeed, the appearance of the 
gluonic operator in the trace of the energy--momentum tensor is a 
reflection of 
the broken scale invariance of QCD. However the chiral symmetry implies 
zero scale dimension for the Goldstone boson fields \cite{Chiv} -- 
otherwise the scale 
transformations would break chiral invariance. 

A closely related result \cite{VZ}, based on the same properties of 
the theory, 
fixes the matrix element of the gluon operator (\ref{3.4}) between the vacuum 
and the two-pion state:
\be
\langle 0|{\tilde{\beta} (g) \over 2g} 
G^{\alpha\beta a}G^a_{\alpha\beta} | \pi^{+}\pi^{-} \rangle = q^2,
\ee
where $q^2$ is the square of the dipion invariant mass. This result leads 
to the suppression of low-mass dipions in the $\psi'\to J/\psi + \pi\pi$, 
$\Upsilon' \to \Upsilon + \pi\pi$ transitions, which is confirmed 
experimentally. (See \cite{cur} for the early applications of the current 
algebra to hadronic cascade transitions and \cite{CN} for a theoretical update 
on the subject.)

It is interesting to note that the decoupling of soft pions from a heavy 
isoscalar target in fact follows from the results of Weinberg \cite{SW},  
obtained in 1966 on the basis of current algebra. 
The formula for the scattering length in the soft pion interaction 
with a heavy target, 
derived in ref. \cite{SW}, reads
\be
a_{T} = -L \left({{1+M_{\pi}}\over {M_{t}}}\right)^{-1} [T(T+1)-T_{t}(T_{t}+1) 
-2]; \label{Wein}
\ee
where $L=g_{V}^2M_{\pi}/2\pi F_{\pi}^2$ gives a characteristic length scale,
$T_{t}$ and $M_{t}$ are the isospin and the mass of the target, and $T$ 
is the total isospin. Putting $T_{t}=0$ as for quarkonium states, we 
get $T=T_{\pi}=1$; in this case the formula (\ref{Wein}) yields a zero 
scattering length, again implying decoupling of soft pions! 

Our result is therefore just a new and directly based on QCD way of deriving 
the low-energy theorem that had been known already for a long time. 
The derivation based on the operator product expansion shows that this 
theorem can be expected to work well 
when the target is tightly bound; in the current algebra approach this 
condition is translated as the absence of any structure in the 
spectral density of the target excitations in the vicinity of the 
ground-state pole\footnote{I thank M. Chemtob for a useful 
discussion on this topic.} at $M_{t}^2$. Heavy quarkonia provide, perhaps, 
the best example of a hadronic system for which this assumption holds; 
we therefore 
expect the decoupling theorem to be quite accurate in this case.

\subsection{The phase of the scattering amplitude}

The phase of the forward scattering amplitude is an important quantity, 
which in general cannot be calculated in QCD from the first principles. 
Usually one has to rely on the predictions of Regge theory, according 
to which, at high energy, the scattering amplitudes are dominated by 
the Pomeron exchange and are almost purely imaginary.

We now have everything at hand to perform a QCD calculation of the phase 
of the forward quarkonium--hadron scattering amplitude. Apart from the 
pure theoretical interest, this quantity is important for practical 
applications, since, for example, it enters the VMD relation between 
the cross sections of quarkonium scattering and photoproduction, 
governs the nuclear shadowing of quarkonium production \cite{BC}, and determines 
the mass shift of quarkonium states in nuclei and in dense 
hadronic matter \cite{KV, LMS, Bro}.      
 
To compute the phase of the forward scattering amplitude, we shall use  
dispersion 
relations. In doing so, it is important to remember that 
in the limit of zero energy, where the amplitude can be calculated 
in a model-independent 
way, it does not vanish (apart from the case of scattering 
on a Goldstone boson, considered in Sect.3.2) and is purely real. 
To reconstruct the real part of the amplitude from the imaginary one, 
we should therefore make a subtraction at zero energy:
\be
F(\d) = F(0) + {1 \over \pi} \int_{\d_0}^{\infty}d\d'\ {{\rm Im} F(\d') 
\over \d'} 
{2\d^2 \over \d'^2 - \d^2}, \label{3.23}
\ee
where the imaginary part of the forward scattering amplitude is related to 
quarkonium absorption cross section by the optical theorem. 
Using the identity
$$
{1 \over x} = P\left({1 \over x}\right) + i\pi \delta (x),
$$
we can split Eq. (\ref{3.23}) in two equations for the real and the imaginary 
parts of the amplitude. The equation for the imaginary part of course reduces 
to a trivial identity; the equation for the real part is
\be   
{\rm Re} F(\d) = F(0) + {1 \over \pi} P \int_{\d_0}^{\infty}d\d'\ 
{{\rm Im} F(\d') 
\over \d'} 
{2\d^2 \over \d'^2 - \d^2}. \label{3.24}
\ee
The equation (\ref{3.24}) allows a numerical reconstruction of the phase of 
the forward 
scattering amplitude as a function of energy  using the results 
of the previous sections. 
The reconstructed amplitude has a substantial real part up to quite 
high energies; 
this shows that even though the only allowed exchanges are purely gluonic, 
the exchange cannot be adequately described by the Pomeron. In fact, 
the Pomeron exchange at high energy leads to almost completely imaginary 
amplitude, which corresponds to the large number of open inelastic channels. 
In our case, the large binding energy of quarkonium suppresses the 
break-up probability, reducing the number of accessible inelastic final states.
  
Moreover, as was already discussed above, the energy dependence of the 
absorption cross section is different from what could be expected from the 
Pomeron exchange. The reason for this is simple: due to the large binding 
energy of quarkonium, the absorption cross section at moderate energies 
reflects essentially the $x \to 1$ behaviour of the gluon structure function, 
whereas the Pomeron governs the $x \to 0$ region. 

\section{Quarkonium interactions in matter}
\renewcommand{\theequation}{4.\arabic{equation}}
\setcounter{equation}{0}

\subsection{Nuclear matter}

Nuclear matter is the best-studied sample of hadronic matter we have 
at our disposal; it is also the most obvious environment to study the 
properties of quarkonium in external fields. Such a study would provide 
a direct check of the results on quarkonium--nucleon scattering amplitude, 
 which is hardly possible otherwise. Indeed, the only other possibility 
to study quarkonium--nucleon scattering stems from the analyses of the 
\J\ and \U\ photoproduction in the framework of the vector meson dominance 
model. This approach, however, suffers from ambiguities in the off-shell 
continuation of the amplitude, which can be dangerous in the most 
interesting region of low energies where the range of extrapolation is 
rather large. Moreover, the extraction of the absorptive part of the 
amplitude from the experimental data on differential cross section of 
photoproduction requires the 
knowledge of the real-to-imaginary ratio of the amplitude. This latter 
is commonly assumed to be equal to zero, in analogy with the known properties 
of light meson--nucleon scattering. 
Basing on the results of Sect.3, we suspect 
that this assumption is wrong, and for heavy quarkonia the real part 
of the amplitude does not vanish, especially at low energies.     
Therefore to extract the quarkonium--nucleon amplitude in a reliable 
way we should turn to nuclear interactions. 

There seems to be a lot of experimental data on quarkonium interactions 
inside nuclei: both quarkonium production in hadron--nucleus 
collisions and leptoproduction were extensively studied. 
It then looks possible to extract the quarkonium--nucleon amplitude from 
the data, applying the Glauber formalism for the final-state interactions. 
Unfortunately the real situation is not that simple. 
Indeed, the production of a physical quarkonium state requires some finite 
time, which can be estimated from the characteristic virtualities of the 
corresponding Feynman diagrams of the colour-singlet approach \cite{CS}. 
The hadro- production of vector states, for example, requires at least 
three gluons 
involved, of which only two must be hard 
to create the $\bar{Q}Q$ pair. The third one can be very soft (note 
that the amplitude is finite in this limit), and this ``explains" 
the failure of perturbative approach in describing the recent high-energy 
 data on 
quarkonium production. A possible way out is to assign this soft gluon to 
the quarkonium wave function, thus introducing, for example, 
the notion of $|\bar{Q}Qg\rangle$ higher Fock state \cite{Braa}. 
This also helps to understand the phenomenological success of the colour 
evaporation model in explaining the data (see \cite{Gav} for a recent study 
and more references).
The proper lifetime of 
the $|\bar{Q}Qg\rangle$ state (estimated as 
$\simeq (2m_Q \Lambda_{{\rm QCD}})^{-1/2}$ 
in ref. \cite{KS1}) in the nucleus rest frame 
will be sufficient for this state  
to traverse the entire nuclear volume. Therefore the observed nuclear 
attenuation of quarkonium production has nothing to do with the absorption 
of physical quarkonium states. To perform a real measurement of the 
quarkonium nucleon cross section, one has to consider interactions of 
quarkonia which are sufficiently slow inside the nucleus. This requires 
measurements in the negative $x_F$ region, which are hard to perform, 
since slow dileptons are hard to measure. 
There is, however, a way out -- one can perform a so-called inverse kinematics 
experiment, in which the nuclear beam is incident on a hydrogen target 
\cite{KS,KS4}. 
In this set-up, the quarkonium states, which are slow inside the nucleus, 
become fast in the lab; they therefore decay into fast dileptons, which are 
easy to detect experimentally. Such an experiment has become feasible with 
an advent of a lead beam at the CERN SPS. It can provide the first measurement 
of quarkonium--nucleon absorption cross section. 

There is another interesting issue related to the interaction of low-energy 
quarkonia inside the nuclear matter. In sect.3, we have found that the 
quarkonium-- nucleon elastic scattering amplitude at 
low energies has to be real 
and correspond to attraction. The quarkonium--nucleus scattering amplitude 
should be constructed as a multiple scattering series. Normally, 
in hadron--nucleus interactions the series converges quite rapidly due to 
the large imaginary part of the elementary scattering amplitude, and the 
first term (``impulse approximation") provides a good description of 
the hadron--nucleus scattering amplitude. The smallness of the imaginary part 
of the quarkonium--nucleon amplitude at low energies however changes the 
situation drastically, and large collective effects are to be expected 
\cite{Can}. 
The most spectacular phenomenon would be the formation of a nuclear bound 
state of quarkonium \cite{Bro, Was, KV, LMS, Was, DK, Can}. 
The characteristic experimental signature of such a state 
would be a shift downwards of the peak in the dilepton spectrum at large 
rapidities (corresponding to small relative velocities of quarkonium and 
residual nucleus).     

\subsection{Hadron gas}

Let us consider first an ideal gas
of pions. Their momentum distribution
is thermal, i.e. for temperatures not too low it is given by
${\rm exp}(-E_{\pi}/T) \simeq {\rm exp}(-p_{\pi}/T)$. Hence the average
momentum of a pion in this medium is $\langle p_{\pi} \rangle = 3T$. The
distribution of gluons within a pion is 
rather soft; the quark counting rules imply that at large $x$ the 
structure functions should decrease at least as fast as $g(x) \sim (1-x)^3$.
(The small-$x$ behaviour does not affect our considerations here). 
As a consequence, the average momentum of a gluon
in confined matter is given by
\be
\langle p_g \rangle_{\rm conf} \leq {1 \over 5} \langle p_{\pi} \rangle =
{3 \over 5} T. \label{av}
\ee
Hence in a medium of temperature $T \simeq 0.2$ GeV, the average gluon
momentum is around 0.1 GeV. Since this is far too small for the break-up 
of tightly bound quarkonium states, a confined pion gas is not effective 
in quarkonium suppression \cite{KS,HS}. 
This statement is confirmed by explicit 
calculations of the thermally-averaged  cross sections of $J/\psi$ 
and $\Upsilon$ absorption in a hadron gas \cite{HS}.

These calculations are based on the formalism described in the previous 
Sections and become exact in the heavy quark limit.  
Nevertheless, in view of the finite charm quark mass, it makes sense
to ask if this formalism correctly describes
\J~interactions with light hadrons. 
Non-perturbative corrections to the \J\ dissociation were analysed, 
in a semi-classical approach, in \cite{Larry}. In this approach, 
the dominant non-perturbative processes leading to the \J\ break-up are 
tunnelling and direct thermal activation to the continuum. 
The rates of these processes were calculated in ref. \cite{Larry} in a 
largely model-independent way in terms of one phenomenological 
parameter $L$ --  
the distance at which charm quarks couple to light quarks and form open-charm 
hadrons. For reasonable values of this parameter $L\leq 1$ fm, neither 
of the considered mechanisms leads to a sufficiently large dissociation to 
explain the experimentally observed suppression of \J . 
More work on the non-perturbative corrections is needed.

\subsection{Deconfined matter}
\medskip

The ultimate constituents of matter are evidently always quarks and
gluons. What we want to know is if these quarks and gluons are
confined to hadrons or not. Let us therefore assume that we are given a
macroscopic volume of static strongly interacting matter and have to
analyse its confinement status.

The distribution of gluons in a deconfined medium is
directly thermal, i.e. ${\rm exp}(-p_g/T)$, so that
\be
\langle p_g \rangle_{\rm deconf} = 3T. \label{5.2}
\ee
Hence the average momentum of a gluon in a deconfined medium is five
times higher than in a confined medium (see Eq. (\ref{av})\footnote{We could
equally well
assume matter at a fixed energy density, instead of temperature. This
would lead to gluons, which are approximately
three times harder in case of deconfinement than for confinement.}; 
for $T=0.2$ GeV, it
becomes 0.6 GeV. An immediate consequence of deconfinement is thus
a considerable hardening of the gluon momentum distribution \cite{KS,HS}. 
Although we
have here presented the argument for massless pions as hadrons, it
remains essentially unchanged for heavier mesons ($\rho/\omega$) or
nucleons, where one can use a non-relativistic thermal distribution for
temperatures up to about 0.5 GeV.
We thus have to find a way to detect such a hardening of the gluon
distribution in deconfined matter.
\par
The lowest charmonium state \J~provides an ideal probe for
this. It is very small, with a radius $r_{\psi}\simeq 0.2$ fm $\ll
\Lambda^{-1}_{QCD}$, so that \J~interactions with the conventional light
quark hadrons probe the short-distance features, the parton
infra structure, of the latter. It is strongly bound, with a
binding energy $\epsilon_{\psi} \simeq 0.65$ GeV $\gg \Lambda_{{\rm QCD}}$;
hence it
can be broken up only by hard partons. Since it shares no quarks or
antiquarks with pions or nucleons, the dominant perturbative interaction
for such a break-up is the exchange of a hard gluon, and this was the
basis of the short-distance QCD calculations presented in Section 2.
\par
We thus see qualitatively how a deconfinement test can be carried out.
If we put a \J~into matter at a temperature $T=0.2$ GeV, then
\begin{itemize}
\item{if the matter is confined, $\langle p_g \rangle_{\rm conf}
\simeq 0.1$ GeV, which is too soft to resolve the \J~as a
$\c$ bound state and much less than the binding energy
$\epsilon_{\psi}$, so that the \J~survives;}
\item{if the matter is deconfined, $\langle p_g \rangle_{\rm deconf}
\simeq$ 0.6 GeV, which (with some spread in the momentum distribution)
is hard enough to resolve the \J~and to break the binding, so that the
\J~will disappear.}
\end{itemize}
\par\noindent
The latter part of our result is in accordance with the mentioned prediction
that the formation of a QGP should lead to a \J~suppression
\cite{MS,HS}. There it was argued that in a QGP, colour screening
would prevent any
resonance binding between the perturbatively produced $c$ and ${\bar
c}$, allowing the heavy quarks to separate. At the hadronization point
of the medium, they would then be too far apart to bind to a \J~and
would therefore form a $D$ and a $\bar D$. Although the details of such
a picture agreed well with the observed \J~suppression \cite{NA38}, it
seemed possible to obtain a similar suppression by absorption in a
purely hadronic medium.
Taking into account the partonic substructure of such hadronic
break-up processes, we now see that this is in fact not possible for
hadrons of reasonable thermal momentum. Our picture thus not only
provides a dynamical basis for \J~suppression
by colour screening, but also indicates that in fact additional suppression 
of physical  \J~
in dense matter will occur {\sl if and only if} there is deconfinement.
We note, however, that the dynamical approach to \J~suppression does not
require a thermal equilibrium of the interacting gluons, so that it
will remain applicable even in deconfined pre-equilibrium stages.

In Section 2.4 we had obtained the cross section for the dissociation of
a tightly bound quarkonium by an incident light hadron. Equation\ (2.37) 
can be 
equivalently obtained \cite{Peskin,HS} by convolution of the inelastic 
gluon--charmonium
cross section with the gluon distribution in the light hadron. The
gluon--quarkonium cross section itself is given by
\be
\sigma_{g\Phi}(k) = {2\pi \over 3} \left( {32 \over 3} \right)^2 \left(
{m_Q \over \e_0} \right)^{1/2} {1 \over m^2_Q} {(k/\e_0 - 1)^{3/2}
\over (k/\e_0)^5 }, \label{5.4}
\ee
with $k$ denoting the momentum of the gluon incident on a stationary
quarkonium. The resulting break up cross section
for gluon--\J~and gluon--\U~interactions as function of the gluon
momentum are broadly peaked in the range $0.7 \leq
k \leq 1.7$ GeV for the \J, with a maximum value of about 3 mb, and in
the range $1.2 \leq k \leq 2.2$ GeV for the \U, with a maximum of about
0.45 mb. The corresponding cross sections
for incident pions (note that now $k=3$ in Eq.\ (4.2)), with high-energy
values of 3 mb and 0.5 mb for \J~and \U, respectively, are 
negligible up to momenta of around 4 GeV
for
the \J~and 7 GeV for the \U. These results thus provide the basis for the
claim that in matter temperature $T\leq 0.5$ GeV, gluons of thermal
momentum can break up charmonia, while hadrons cannot.
We note here that, just as in the photoelectric dissociation of atoms,
the break-up is most effective when the momentum of the gluon is
somewhat above the binding energy. Gluons of lower momenta can neither
resolve the constituents in the bound state nor raise them up to the
continuum; on the other hand, those of much higher
momenta just pass through it.
\par
To illustrate this more explicitly, we calculate the break-up cross
section for the \J~as a function of the temperature $T$ of an ideal QGP.
Using Eq. (\ref{5.4}) with $m_c=1.5$ GeV and the \J~binding energy of 0.64
GeV, we then get
\be
\sigma_{g\j}(T) \simeq 65~{\rm mb} \times { \int_{\e_0}^{\infty} dk~k^2
e^{-k/T}
(k/\e_o-1)^{3/2}
(k/\e_0)^5 \over \int_{\e_0}^{\infty} dk~k^2 e^{-k/T}}. \label{5.5}
\ee
The effective cross section for break-up in the temperature range
$0.2 \leq T \leq 0.5$ GeV is about 1.2 mb. It is this value that will
determine the suppression of the (pure $1S$) \J~in a deconfined medium.
\par
In nuclear collisions, the medium is certainly not static. To perform a 
calculation of the quarkonium production in such conditions, one therefore 
needs to evoke some model to describe the collision 
dynamics. This was attempted in ref. \cite{HS}, where the idealized case 
of an isentropic longitudinal expansion of a thermally equilibrated medium 
was considered. A more realistic model considering the quarkonium 
interactions in equilibrating parton gas was considered in ref. \cite{XW}. 
A discussion of the phenomenology of quarkonium suppression based on 
different models of nuclear collision dynamics lies beyond the scope 
of this review. However the work in this direction is certainly necessary 
to understand the experimental data.

\section{Discussion and Outlook}
\medskip
\par
Because of the small size and the large binding energy of the lowest
quarkonium states, their interaction with light hadrons is calculable in
short-distance QCD. They can interact in leading
order only through the exchange of a hard gluon, and the gluon
distribution in the light hadrons is known to be very soft. The resulting
prediction is a cross section that rises very slowly from threshold
to its high-energy value, suppressing strongly any break-up of
quarkonium ground states by slow mesons or nucleons.
\par
Low-energy theorems of QCD, based on the concepts of spontaneously broken 
scale and chiral symmetries, allow us to calculate the amplitudes of 
quarkonium interactions with slow hadrons in a model-independent way. 
It can be shown, in particular, that tightly bound quarkonium states 
decouple from soft pions.

As a consequence of these results, confined matter at meaningful temperatures 
 becomes
transparent to \J's and \U's. The momentum of deconfined thermal gluons, on the
other hand, is large enough to give rise to effective \J~dissociation;
such dissociation can occur also by deconfined gluons which are not in
equilibrium.
Strongly interacting matter thus leads to \J~suppression if and only if
it is deconfined. The loosely bound \P~can be broken up in both confined
and deconfined matter, though presumably more in a deconfined medium.

In hadron--nucleus and nucleus--nucleus collisions, the observed suppression 
results not only from the dissociation of physical quarkonium states, 
but also from the nuclear attenuation of the quarkonium production process. 
The recent data on quarkonium production in high-energy hadronic collisions 
suggest the dominance of intermediate higher Fock states, such as 
$|\bar{Q}Qg\rangle$, in this process \cite{Braa}. 
The nuclear attenuation of such 
states was recently considered in ref. \cite{Comp}. Since the $\bar{Q}Q$ pair 
in such states is in a colour-octet state, where gluonic exchanges are 
repulsive (see eq. (2.4)), the $|\bar{Q}Qg\rangle$ states can be easily 
dissociated. 
 It was found \cite{Comp} that the  suppression 
of \J\ and \U\ 
states  at present observed in $h-A$ and $A-A$ collisions can be 
completely accounted for 
in terms of the $|\bar{Q}Qg\rangle$ state absorption in confined nuclear 
matter. This result provides the basis for the phenomenologically successful 
Gerschel--H\"uffner fit \cite{GH} and is also qualitatively 
consistent with the findings of 
ref. \cite{Cap}.

The equality of \J~and \P~suppression in $h-A$ collisions for $x_F\geq
0$, as well as the size and $x_F$ dependence of the observed effect, are
in full
accord with the passage of a $|\bar{Q}Qg\rangle$ state through nuclear 
matter. The
equality of \J~and \P~suppression and the observed $x_F$ dependence
are in clear disagreement with any description based on
the absorption of fully formed physical charmonium states in nuclear matter.

There is a lack of data for charmonium production in a kinematic regime
in which fully formed \J's could interact with nuclear matter. Such data
could be obtained by experiments using the Pb-beam incident on a light
target \cite{KS4}.
\par
The observed additional \P~suppression in nucleus--nucleus
interactions \cite{NA38-new} indicates the presence of confined hadronic 
matter at later stage of the collision, when the physical quarkonium 
states are formed. According to the results reviewed here, the presence of 
confined matter does not lead to additional suppression of tightly bound 
\J\ states, but can result in the strong additional suppression of loosely 
bound \P's. It would be interesting to see if the 
Pb-beam data forthcoming from the 
CERN SPS show an additional suppression of \J's. If found, 
either at the SPS or at future experiments at RHIC and the LHC, 
this suppression 
can signal the presence of collective partonic effects. 

\vskip0.8cm
{\bf Acknowledgements}
\vskip0.2cm
I am grateful to Profs. A. Di Giacomo and D. Diakonov for their invitation 
to this excellent School.  

I wish to thank H. Satz for an enjoyable 
collaboration on the topics considered 
in this review. The results on the non-perturbative interactions of quarkonium 
were obtained in collaboration with L. McLerran, whom I also thank for 
valuable comments and discussions. 
I am grateful to my other collaborators on quarkonium production studies: 
R. Gavai, G. Schuler, K. Sridhar and R. Vogt. 
I thank X.-N. Wang and X.-M. Xu, 
together with whom a calculation of \J\ suppression in an equilibrating parton 
plasma was carried out. The financial support of the German Research Ministry
(BMBW) under contract 06 BI 721 is gratefully acknowledged.

% stop


\newpage
\vskip1cm
\begin{thebibliography}
\medskip
\bibitem{dec}
{T. Appelquist and H.D. Politzer, Phys. Rev. Lett. 34 (1975) 43;\\
A. de R\'ujula and S.L. Glashow, Phys. Rev. Lett. 34 (1975) 46;\\
R. Barbieri, R. Gatto and R. K\"ogerler, \PL B 60 (1976) 183;\\
V.A. Novikov, L.B. Okun, M.A. Shifman, A.I. Vainshtein, M.B. Voloshin and 
V.I. Zakharov, Phys. Rep. 78 (1978) 1.}
\bibitem{Ger}
{G.A. Schuler, CERN-TH.7170/94; Phys. Rep., to appear.}
\bibitem{SVZ} 
{A. I. Vainshtein, V.I. Zakharov and M.A. Shifman, JETP Lett. 27 (1978) 60;\\
M.A. Shifman, A.I. Vainshtein, M.B. Voloshin and V.I. Zakharov, \PL B 77 (1978) 
80;\\
 M.A. Shifman, A.I. Vainshtein and V.I. Zakharov, \NP B 147 (1979) 385, 448.} 
\bibitem{MS}
{T. Matsui and H. Satz, \PL B 178 (1986) 416.}
\bibitem{NA38}
{NA38 Collaboration, C. Baglin et al., \PL B 220 (1989) 471; B 251 (1990) 
 465, 472; B225 (1991) 459.}
\bibitem{Blaizot}
{A. Capella, J.A. Casado, C. Pajares, A.V. Ramallo and J. Tr\^an Thanh V\^an, 
\PL B 206 (1988) 354;\\
J. Ftacnik, P. Lichard and J. Pis\'ut, \PL B 207 (1988) 194;\\
C. Gerschel and J. H\"ufner, \PL B 207 (1988) 253;\\
S. Gavin, M. Gyulassy and A. Jackson, \PL B 207 (1988) 257;\\
J.-P. Blaizot and J.-Y. Ollitrault, \PL B 217 (1989) 392;\\
for a review, see J.-P. Blaizot and J.-Y. Ollitrault, in: 
``Quark-Gluon Plasma", R.C. Hwa (Ed.), World Scientific, Singapore, 1990.}
\bibitem{Got}
{K. Gottfried, \PRL 40 (1978) 598.}
\bibitem{Vol}
{M.B. Voloshin, \NP B 154 (1979) 365;\\
Sov.J.Nucl.Phys. 36 (1982) 143; 40 (1984) 662.}
\bibitem{Leut}
{H. Leutwyler, \PL B 98 (1981) 447.} 
\bibitem{time}
{D. Gromes, \PL B 115 (1982) 482;\\
V.N. Baier and Yu.F. Pinelis, \PL B 116 (1982) 179;\\
M. Campostrini, A. Di Giacomo and S. Olejnik, \ZP C 31 (1986) 577;\\
H.G. Dosch, \PL B 190 (1987) 177;\\
H.G. Dosch and Yu.A. Simonov, \PL B 205 (1988) 339;\\
Yu.A. Simonov, \NP B 234 (1989) 67;\\
A. Kr\"amer, H.G. Dosch and R.A. Bertlmann, \PL B 223 (1989) 105;\\
Yu.A. Simonov, S. Titard and F.J. Yndurain, \PL B 354 (1995) 435.}
\bibitem{Peskin}
{M. E. Peskin, \NP B156 (1979) 365.}
\bibitem{Bhanot}{G. Bhanot and M. E. Peskin, \NP B156 (1979) 391.}
\bibitem{Kaidalov}{A. Kaidalov, in {\sl QCD and High Energy
Hadronic Interactions}, J. Tr\^an Thanh V\^an (Ed.), Editions Frontieres,
Gif-sur-Yvette, 1993.}
\bibitem{KS}
{D. Kharzeev and H. Satz, \PL B 334 (1994) 155.}
\bibitem{ES}
{E.V. Shuryak, Sov.J.Nucl.Phys. 28 (1978) 408.}
\bibitem{HS}
{D. Kharzeev and H. Satz, CERN-TH/95-117; in: ``Quark-
Gluon Plasma II", R.C. Hwa (Ed.), World Scientific, Singapore, 1995;\\
H. Satz, \NP A 590 (1995) 63c;\\
D. Kharzeev and H. Satz, \NP A 590 (1995) 515c.}
\bibitem{Gross}
{D. J. Gross and F. Wilczek, \PR D8 (1973) 3633
and \PR D9 (1974) 980;\\
H. Georgi and H. D. Politzer, \PR D9 (1974) 416.}
\bibitem{SVZ1}
{M. A. Shifman, A. I. Vainshtein and V. I. Zakharov,
\PL 65B (1976) 255.}
\bibitem{Novikov}
{V. A. Novikov, M. A. Shifman, A. I. Vainshtein and V. I. Zakharov,
\NP B136 (1978) 125.}
\bibitem{Parisi}
{G. Parisi, \PL 43B (1973) 207; 50B (1974) 367.}
\bibitem{Povh}
{J. H\"ufner and B. Povh, \PRL 58 (1987) 1612.}
\bibitem{MT}
{F. Karsch, M. T. Mehr and H. Satz, \ZP C 37 (1988) 617.}
\bibitem{JE}
{J. Ellis, \NP B 22 (1970) 478;\\
R.J. Crewther, \PL B 33 (1970) 305; \PRL 28 (1972) 1421;\\
M.S. Chanowitz and J. Ellis, \PL B 40 (1972) 397; \PR D7 (1973) 2490.}
\bibitem{scale}
{J. Collins, A. Duncan and S.D. Joglecar, \PR D16 (1977) 438;\\
N.K. Nielsen, \NP B120 (1977) 212.}
\bibitem{VZ}
{M. Voloshin and V. Zakharov, \PRL 45 (1980) 688.}
\bibitem{KV}
{A.B. Kaidalov and P.E. Volkovitsky, \PRL 69 (1992) 3155.}
\bibitem{LMS}
{M. Luke, A.V. Manohar and M.J. Savage, \PL B 288 (1992) 355.}
\bibitem{SVZ2}
{M.A. Shifman, A.I. Vainshtein and V.I. Zakharov, \PL B 78 (1978) 443.}
\bibitem{GLS}
{J. Gasser, H. Leutwyler and M.E. Sainio, \PL B 253 (1991) 252.}
\bibitem{Bro}
{S.J. Brodsky, I. Schmidt and G.F. de Teramond, \PRL 64 (1990) 1011.}
\bibitem{Chiv}
{R.S. Chivukula, A. Cohen, H. Georgi, B. Grinstein and A.V. Manohar, 
Ann. Phys. 192 (1989) 93.}
\bibitem{cur}
{L.S. Brown and R. Cahn, \PRL 35 (1975) 1;\\
M.B. Voloshin, JETP Lett. 21 (1975) 733;\\
R. Cahn, \PR D 12 (1975) 3559.}
\bibitem{CN}
{M. Chemtob and H. Navelet, \PR D 41 (1990) 2187; Annales de Physique 15 
(1990) 243.}
\bibitem{SW}
{S. Weinberg, \PRL 17 (1966) 616.}
\bibitem{BC}
{D. Kharzeev and H. Satz, \PL B 327 (1994) 361;\\
A. Bialas and W. Czyz, \PL B 328 (1994) 172;\\
for a discussion of quantum effects in shadowing, see also
S.J. Brodsky and H.J. Lu, \PRL 64 (1990) 1342.}
\bibitem{CS}
{C.H. Chang, \NP B 172 (1980) 425;\\
E.L. Berger and D. Jones, \PR D 23 (1981) 1521;\\
R. Baier and R. R\"uckl, \PL B 102 (1981) 364; \ZP C 19 (1983) 251.}
\bibitem{data}
{The UA1 Collaboration, C. Albajar et al., \PL B 256 (1991) 112;\\
The CDF Collaboration, F. Abe et al., \PRL 69 (1992) 3704; 71 (1993) 2537;
FERMILAB-PUB-95/271-E;\\
The D0 Collaboration, S. Abachi et al., FERMILAB-CONF-95-205-E;\\  
FERMILAB-CONF-95-361-E.}  
\bibitem{Braa}
{G.T. Bodwin, E. Braaten and G.P. Lepage, \PR D 51 (1995) 1125;\\
E. Braaten and S. Fleming, \PRL 74 (1995) 3327;\\
M. Cacciari, M. Greco, M.L. Mangano and A. Petrelli, \PL B 356 (1995) 553;\\
P. Cho and A.K. Leibovich, CALT-68-1988 (1995); CALT-68-2026 (1995);\\
E. Braaten, NUHEP-TH-95-11 (1995);\\
M.L. Mangano, CERN-TH/95-190.}
\bibitem{Gav} 
{R. Gavai, D. Kharzeev, H. Satz, G. Schuler, K. Sridhar and R. Vogt, 
Int. J. Mod. Phys. A 10 (1995) 3043.}
\bibitem{KS1}
{D. Kharzeev and H. Satz, \ZP C 60 (1993) 389.}
\bibitem{KS4}
{D. Kharzeev and H. Satz, \PL B 356 (1995) 365.}
\bibitem{Can}
{F. Cannata, D. Kharzeev and F. Piccinini, \PR C 49 (1994) 2798.} 
\bibitem{Was}
{D.A. Wasson, \PRL 67 (1991) 2237.}
\bibitem{DK}
{D. Kharzeev, \NP A 558 (1993) 331c.}
\bibitem{Larry}
{D. Kharzeev, L. McLerran and H. Satz, \PL B 356 (1995) 349.}
\bibitem{XW}
{X.-M. Xu, D. Kharzeev, H. Satz and X.-N. Wang, LBL-37985, 
CERN-TH/95-304 (1995).}
\bibitem{Comp}
{D. Kharzeev and H. Satz, CERN-TH/95-214 (1995); \PL B, 
in press.}
\bibitem{GH}
{C. Gerschel and J. H\"ufner, \ZP C 56 (1992) 171.}
\bibitem{Cap}
{A. Capella, C. Merino, J. Tr\^an Thanh V\^an, C. Pajares and 
A. Ramallo, \PL B 243 (1990) 144; \\
G. Piller, J. Mutzbauer and W. Weise, \ZP A 343 (1992) 247; \NP A 560 (1993) 
437;\\
K. Boreskov, A. Capella, A. Kaidalov and J. Tr\^an Thanh V\^an, 
\PR D 47 (1993) 
919;\\
for a related discussion, see also U. Heinz and R. Wittmann,
Z. Phys. C 59 (1993) 77;\\
M. Ga\'zdzicki and S. Mr\'owczy\'nski, Z. Phys. C 49 (1991) 569.}
\bibitem{NA38-new}
{The NA38 Collaboration, B. Ronceux et al., \PL B 345 (1995) 
617.}

\end{thebibliography}
\end{document}